\documentclass[prd,nofootinbib]{revtex4}
\usepackage{amsmath}
\usepackage{graphicx}
\usepackage{amssymb}

\begin{document}

\title{Renormalized 2PN spin contributions to the accumulated orbital phase\\
for LISA sources}
\author{L\'{a}szl\'{o} \'{A}rp\'{a}d Gergely$^{1,2,3\dag }$, Bal\'{a}zs Mik%
\'{o}czi$^{1,2,4\ddag }$}
\affiliation{$^{1}$ Department of Theoretical Physics, University of Szeged, Tisza Lajos
krt. 84-86, Szeged 6720, Hungary\\
$^{2}$ Department of Experimental Physics, University of Szeged, D\'{o}m t%
\'{e}r 9, Szeged 6720, Hungary\\
$^{3}$ Department of Applied Science, London South Bank University, 103
Borough Road, London SE1 0AA, UK\\
$^{4}$ KFKI Research Institute for Particle and Nuclear Physics, Budapest
114, P.O.Box 49, H-1525 Hungary \\
$^{\dag }$gergely@physx.u-szeged.hu, $^{\ddag }$mikoczi@rmki.kfki.hu; }

\begin{abstract}
We give here a new third post-Newtonisn (3PN) spin-spin contribution (in the
PN parameter $\varepsilon $) to the accumulated orbital phase of a compact
binary, arising from the spin-orbit precessional motion of the spins. In the
equal mass case this contribution vanishes, but LISA sources of merging
supermassive binary black holes have typically a mass ratio of 1:10. For
such non-equal masses this 3PN correction is periodic in time, with period
approximately $\varepsilon ^{-1}$ times larger than the period of
gravitational waves. We derive a \textit{renormalized} and simpler
expression of the spin-spin coefficient\ at 2PN, as an average over the
time-scale of this period of the combined 2PN and 3PN contribution. We also
find that for LISA sources the quadrupole-monopole contribution to the phase
dominates over the spin-spin contribution, while the self-spin contribution
is negligible even for the dominant spin. Finally we define a renormalized
total spin coefficient $\overline{\sigma }$ to be employed in the search for
gravitational waves emitted by LISA sources.
\end{abstract}

\date{\today }
\maketitle

\section{Introduction}

Gravitational waves are generated by time-varying quadrupolar (or higher
multipolar) mass (energy) configurations in general relativity. Although a
convincing indirect proof for the existence of the theoretically predicted
gravitational waves is provided by binary pulsars \cite{HTGW}, the most
well-known being the Hulse-Taylor pulsar \cite{HT} (others are also known 
\cite{newp1},\cite{newp2}) a major effort is undergoing nowadays for the
first direct detection of gravitational waves by the most sensitive
Earth-based interferometric gravitational wave detectors LIGO \cite{LIGO}
and VIRGO \cite{VIRGO}.

These advanced instruments are able to detect in the frequency range $%
10-1000 $ Hz. Among their most likely sources are compact binary systems
composed of neutron stars / stellar size black holes. In order to simplify
the route towards the first detection, some of the relevant astrophysical
details of the system are commonly omitted, like the spins and quadrupole
moments. The detection effort concentrates on the signal emitted by point
masses on quasi-circular orbit. With this, only the mass ratio should be
monitored, while the inclusion of the astrophysical characteristics would
induce many new parameters (additional six, for example, by the two spins).
However the ambitious future aim of a gravitational wave based astrophysics
would certainly require to include the spins and quadrupole moments of the
binary components in the data analysis.

Supermassive black holes (SMBHs) of mass $3\times 10^{6}\div 3\times 10^{9}$
solar masses are known to exist in the center of the majority of galaxies 
\cite{Kormendy}, \cite{Sanders}, \cite{Faber}, including ours. When galaxies
merge, and the separation of the SMBHs is of a few parsecs, the SMBH binary
enters a dissipative regime dominated by dynamical friction (an interaction
of the SMBHs with the already merged stellar environment). As the relative
separation further shrinks, this is followed by the gravitational radiation
dominated dissipative regime, starting at approximately one hundredth of a
parsec.

During dynamical friction some of the orbital angular momentum of the binary
black hole system is transferred to the stellar environment. In the process
the stellar population at the poles of the system is ejected and a torus is
formed \cite{ZierBiermann}, \cite{Zier}. There had been a major worry, that
the loss-cone mechanism for feeding stars into orbits that intersect the
binary black holes is too slow and the SMBHs will stall in their approach to
each other (the last parsec problem; for a review see \cite{Merritt}).
However recently more than one mechanism was proposed, which lead to a
further approach of the SMBHs: (1) direct interaction with the surrounding
stars slightly further outside \cite{Zier}; (2) relaxation processes due to
cloud / star - star interactions \cite{Alexander}, both repopulating the
stellar orbits in the center of the galaxy. In a series of papers it has
been recently shown that (3) even in the absence of two-body relaxation or
gas dynamical processes, unequal mass and / or eccentric binaries with the
mass larger than $10^{5}$ solar masses can shrink to the gravitational wave
emission regime of one hundredth of a parsec in less than a Hubble time due
to the binary orbital decay by three-body interactions in the
gravitationally-bound stellar cusps \cite{Sesana}. Alternatively, by
considering (4) the three accretion disks: one around each black hole and a
third one, which is circumbinary, it has been shown in \cite{Hayasaki} that
the balance of the orbital angular momentum changing processes (the
circumbinary disk removes orbital angular momentum from the binary via the
binary-disk resonant interaction, however the mass transfer to each
individual black hole adds orbital angular momentum to the binary) is such
that these binaries will merge within a Hubble time driven by this mechanism.

In consequence by one mechanism or another the SMBHs would be able to
approach each other to distances smaller than approximately one hundredth
parsecs, when the gravitational radiation becomes the dominant dissipative
effect. The gravitational waves in the frequency range $10^{-4}-10^{-1}$ Hz
emitted by such a system of coalescing SMBHs will be captured by the
forthcoming LISA space mission \cite{LISA}.

There are a few important comments to make on these SMBH LISA sources.

(a1) The mass distribution of the galactic central SMBHs shows a broken
powerlaw \cite{PressSchechter}, \cite{WilsonColbert}, confirmed by most
recent work \cite{Lauer} and a recent observational survey \cite{Ferrarese}.

(a2) Let $\nu =m_{2}/{m}_{1}$ be the mass ratio of the binary with masses $%
m_{2}\leq m_{1}$. Thus $0\leq \nu \leq 1$. Next we define $m=m_{1}+m_{2}$ as
the total mass, and $\mu =m_{1}m_{2}/m$ as the reduced mass, ranging between 
$\nu $ for $\nu \rightarrow 0$ to $1/4$ for equal masses. Further we
introduce the symmetric mass ratio $\eta =\mu /m=\left( 2+\nu +\nu
^{-1}\right) ^{-1}$ (for small mass ratios $\eta \rightarrow \nu $). The
probability for a specific mass ratio for SMBH encounters can be derived as
an integral over the black hole mass distribution, folded with the rate to
actually merge (proportional to the capture cross section and the relative
velocity for two galaxies). By this method it has been estimated in \cite%
{spinflip} that the most likely mass ratios are in the range $\nu \in \left(
1/30,~1/3\right) $. Therefore the most typical mass ratio for LISA sources
would be $\nu =10^{-1}$, to be considered in this paper.

(b1) The spin of individual black holes has a quite high value. As the
surrounding matter is accreted into the hole, the angular moment transfer
spins up the black hole, close to the maximally allowed value. By taking
into account the angular momentum transfer from accreting matter alone, the
dimensionless spin parameter $\chi =S/m^{2}$ (in units $c=1$\thinspace $=G$)
grows to the maximally allowed value of $1$ even if the initial state of the
black hole was non-rotating \cite{Baardeen}. By taking into account the
energy input of the in-falling (horizon-crossing) photons emitted from the
steady-state thin accretion disk, a torque counteracting to the one due to
mass accretion has been found \cite{PageThorne}, reducing the limiting value
of the dimensionless spin parameter to $0.9982$. The description of the
process can be further enhanced by (i) including the contribution of the
open or closed magnetic field lines in the vicinity of the disk \cite%
{MagneticAccretion}, (ii) discussing a symbiotic system (introduced in \cite%
{JetDiskSymbiosys}) of a rotating black hole, a magnetized thin accretion
disk and jets in the magnetosphere of the hole \cite{KovacsBiermannGergely}.
With any of these refinements, the end state of the black hole will stay
very close to the extremal Kerr limit.

(b2) In consequence the spin of the individual components of a binary black
hole system, which is formed typically by capture events (even for stellar
mass black holes the emergence of binary systems by stellar evolution is
quite unlikely) will be also high.\footnote{%
How large is the final spin after the merger occurs, is still the object of
debate. Recent numerical work on the merger of two black holes with spin has
predicted (beside recoil effects arising for particular configurations \cite%
{recoil}, interesting in their own), the final spin state of the merged
black hole. Extrapolating from the particular configurations analyzed, an
analytical formula was also proposed for the final spin \cite{Rezzolla}.
This formula has been employed in the analysis of the final spin as arising
from a combination of accretion and merger sequence \cite{BertiVolontieri}.
Alternative analyses of the final spin in mergers have been performed by
analytical or numerical methods, yielding various predictions for the values
of the final spin \cite{finalspin}.}

\textit{We conclude from these that any reasonable analysis of the
gravitational waves produced by central galactic SMBH mergers should take
into account both the individual spins} $S_{\mathbf{i}}$ \textit{of the
black holes and their typical mass ratio. }

In the inspiral phase of the merger, which can be described by
post-Newtonian (PN) techniques and lasts until the innermost stable orbit
(ISO) is reached, the spins and quadrupole moments induce a
quasi-precessional evolution of the spin vectors \cite{BOC}. The leading
contributions come from the SO precessions at 1.5PN order. The interaction
of the individual spins lead to a 2PN spin-spin (SS) quasi-precessional
contribution to the dynamics. Similar 2PN quasi-precessional contributions
arise, when one binary component is regarded as a monopole moving in the
quadrupolar field of the other (quadrupole-monopole, QM contribution). This
angular evolution of the spins is decoupled from the radial motion, which
was completely integrated in terms of a generalized Kepler equation \cite%
{Kepler}. The inclusion of the angular evolution (spin precessions) into the
description besides increasing the accuracy, also breaks many degeneracies
among parameters, greatly improving (relative to previous estimates) how
well LISA will be able to measure the masses, the locations of binaries on
the sky, and the distance to the source \cite{LH}. The gravitational
waveforms for spinning compact binaries with precession effects through
1.5PN order were also recently analyzed using spin-weighted -2 spherical
harmonics, and applied to spinning, non-precessing binaries \cite{ABFO}.

A careful and detailed discussion, following the assumptions of Ref. \cite%
{ACST}, by taking into account both the SO precession and the dissipative
effect of gravitational radiation (averaged over one radial period) for the
above typical mass ratio has already resulted in two achievements \cite%
{spinflip}. The first is a quantitative description of the spin-flip
process, visible in the X-shaped radio galaxies. Secondly, the precessional
phase of the merger of two black holes, occurring prior to the spin-flip,
was suggested to be the source of the superdisk, visible in radio galaxies.
The precessing jet acts as a super-wind separating the two radio lobes in
the final stages of the merger. According to this model such radio galaxies
are candidates for subsequent SMBH mergers \cite{spinflip}.

The actual detection of gravitational waves requires the precise knowledge
of the waveforms. This is determined both by the relative geometry of the
sources and detectors, and by the evolution of the sources. In the latter,
the accumulated orbital phase $\phi $ plays a central role, as any effect
that causes the template to differ from the actual signal by one cycle over
the 500 to 16,000 accumulated cycles in the sensitive bandwidth will result
in a substantial reduction in the signal-to-noise ratio \cite{BDWW}.
Therefore the phase of the gravitational waves emitted by a compact binary
is required to high PN accuracy. Various contributions up to 3.5 PN orders
are known. Among these, there are important spin and mass quadrupole
contributions, like SO at 1.5PN, also SS, QM and SS-self contributions at
2PN. (The latter arise from the self-interaction of the spins \cite{MVG}.)
We note here, that alternative definitions of the phase in the spinning case
were introduced (the orbital phase with respect to the ascending node \cite%
{BCV2}) and the rigorous waveform containing SO contributions has been also
computed \cite{Matyi}.

Besides possible quadrupolar irregularities in the mass distribution of the
binary components (which are however quite unlikely due to the strong
self-gravity of these objects yielding rapidly to a spherical
configuration), and the mutual quadrupolar deformation of them (shown to be
quite small in \cite{BlanchetLivRev}), the proper rotation is the effect
generating a considerable quadrupole moment for each black hole. The
quadrupolar deformation can be characterized by the coefficients $p_{i}=$ $%
Q_{i}/m_{i}m^{2}$, with $Q_{i}$ the quadrupole scalar. If the quadrupole
moment originates entirely in the rotation of the compact object (what we
shall assume in what follows), then $Q_{i}\simeq -\alpha \chi
_{i}^{2}m_{i}^{3}$, with the parameter $\alpha \in \left( 4,~8\right) $ for
neutron stars, depending on their equation of state. Stiffer equations of
state give larger values of $a$, while $\alpha =1$ for rotating black holes 
\cite{Poisson}. Here $\chi _{i}=S_{i}/m_{i}^{2}$ are the dimensionless
rotation parameters and the negative sign is because the rotating compact
object is centrifugally flattened, becoming an oblate spheroid.

In what follows, we will consider maximally rotating SMBHs ($\chi
_{i}\approx 1$, thus $S_{i}\approx m_{i}^{2}$), with their quadrupole moment
entirely determined by the rotation. Therefore $%
p_{i}=-S_{i}^{2}/m_{i}^{2}m^{2}\approx -\left( m_{i}/m\right) ^{2}$, thus 
\begin{eqnarray}
p_{1} &\approx &-\left( 1+\nu \right) ^{-2}~,  \label{p1} \\
p_{2} &\approx &-\left( 1+\nu ^{-1}\right) ^{-2}.  \label{p2}
\end{eqnarray}%
The ratio of the spins of the compact objects (characterized by $%
R_{i}\approx m_{i}$ and evaluated for comparable and high rotation
velocities $V_{1}\approx V_{2}\approx 1$ according to our assumption of
maximal rotation) can be expressed as 
\begin{equation}
\frac{S_{2}}{S_{1}}\approx \left( \frac{m_{2}}{m_{1}}\right) ^{2}=\nu ^{2}~,
\label{S1S2}
\end{equation}%
while in the same approximation the ratio of the spins with the magnitude $L$
of orbital angular momentum $\mathbf{L}$ becomes%
\begin{eqnarray}
\frac{S_{2}}{L} &\approx &\frac{m_{2}^{2}V_{2}}{\mu rv}=\left( \frac{m}{r}%
\right) \left( \frac{1}{v}\right) V_{2}\frac{m_{2}}{m_{1}}\approx
\varepsilon ^{1/2}\nu ~,  \label{S2L} \\
\frac{S_{1}}{L} &=&\frac{S_{2}}{L}\frac{S_{1}}{S_{2}}\approx \varepsilon
^{1/2}\nu ^{-1}~,  \label{S1L}
\end{eqnarray}%
where we have employed the definition of the PN parameter $\varepsilon
\approx m/r\approx v^{2}$. As remarked in \cite{spinflip}, in case of a
small $\nu \approx \eta $ the ratio $S_{2}/L$ is shifted towards even
smaller values (therefore $S_{2}\ll L$ during all stages of the inspiral),
however the ratio of the spin of the dominant compact object to the
magnitude of the orbital angular momentum can change drastically. Indeed, it
is determined by the relative magnitude of the small parameters $\varepsilon 
$ and $\nu $. As $\varepsilon $ increases during the inspiral, whenever $\nu 
$ falls in the range of $\varepsilon ^{1/2}$, an $S_{1}\approx L$ epoch is
reached, which follows the initial epoch with $S_{1}<L$ and precedes the
forthcoming $S_{1}>L$ epoch. This happens for the mass ratios $\nu \in
\left( 1/30,~1/3\right) $.

We also note that in previous works we have found useful to choose as some
of the parameters for the new spin degrees of freedom the relative angles $%
\kappa _{i}$ between the spins and orbital angular momentum and the angle $%
\gamma $ between the spins. This is because both the quasi-precessional
equations for the spins and the equations describing the gravitational
radiation losses up to 4.5 PN orders close in these variables, the energy $E$
and magnitude of the orbital angular momentum (averaged over one radial
orbit) $\overline{L}$ \cite{GPV}, \cite{spinspin1}, \cite{spinspin2}, \cite%
{quadrup}.

In this paper we will review the contributions induced by the spins and
quadrupole moments to the accumulated orbital phase of the binary and we
propose modifications in this description. In order to do so, in Section II
we review the existing contributions up to 3.5PN orders. Then we represent
graphically the dependence of the QM contribution on the parameter space $%
\left( \kappa _{1},~\kappa _{2}\right) $. We illustrate that due to the
involved numerical coefficients, the self-spin contribution, although of 2PN
order, is much smaller, than the QM contribution, thus it can be safely
dropped at this accuracy. Both the QM and the SS-self contributions are
discussed for either equal or unequal masses. The dependence of the SS
contribution on the parameter space $\left( \kappa _{1},~\kappa _{2}\right) $
and on the relative spin configuration, expressed in terms of the relative
spin azimuthal angle $\Delta \psi $ is discussed only in the equal mass case
here. (Due to a spherical triangle identity this variable is connected to
the previously mentioned angles $\kappa _{i}$ and $\gamma $.)

In Section III we discuss the consequences of the leading order precession
effects on these coefficients. We show that due to the SO precession the SS
contribution to the phase receives 3PN time-dependent corrections, which
happen to cancel in the equal mass case. At non-equal masses however these
contributions vary periodically, with a period $T_{3PNSS}$ which is $%
\varepsilon ^{-1}$ times longer than the period of the gravitational wave $%
T_{wave}$.

Then in Section IV we look for LISA sources of typical mass ratio $\nu
=10^{-1}$. We represent here the detailed time evolution of the total
spin-spin coefficient on the time scale \thinspace $T_{3PNSS}$ in the
parameter space $\left( \kappa _{1},~\kappa _{2}\right) $. Next we repeat
this for the sum of SS and QM contributions, showing that the QM effects
dominate and the SS effects modulate. We are then able to define a
renormalized SS coefficient at 2PN, which can be considered constant up to
3PN on a time-scale $T_{3PNSS}=\varepsilon ^{-1}T_{wave}$. This coefficient
is simpler than its unnormalized counterpart.

In Section V we compare the number of cycles left\ for typical LISA sources,
considering the following cases: (a) the 2PN accurate spin-spin coefficient,
previously employed in the literature, with the relative spin angles $\gamma
,\kappa _{i}$ being constants at this accuracy (the precessions contributing
only with higher order terms, thus dropped here); (b) our renormalized
spin-spin coefficient (constant up to 3PN), without the precession
equations, which are already taken care of to this accuracy by our
renormalization procedure; (c) the numerical evolution, with time-varying
spin-spin coefficient, found by including the precessions.

Finally in the Concluding Remarks we summarize our findings and define the
renormalized value of the total coefficient encompassing both the QM and SS
contributions, which due to its simplicity, time-independence up to 3PN, and
much better agreement with the numerical results is more suitable to be used
in future LISA data analysis than the presently available expression.

Throughout the paper we use units $G=1=c$.

\section{The phase of gravitational waves to 3.5 PN accuracy}

The integrated orbital phase of gravitational waves has been computed to
high accuracy, with contributions both from general relativistic corrections
to the Keplerian motion and other Newtonian contributions due to the
finite-size of the compact binary components. The respective contributions
were recently summarized in \cite{phi3PN3.5PN}, \cite{MVG}, and \cite{PNSO2}%
. Up to the desired 3.5 PN accuracy the phase can be formally decomposed as:%
\begin{equation}
\phi =\phi _{c}+\phi _{N}+\phi _{1PN}+\phi _{1.5PN}+\phi _{2PN}+\phi
_{2.5PN}+\phi _{3PN}+\phi _{3.5PN}~,  \label{phase}
\end{equation}%
where $\phi _{c}$ is a constant, and the lower order contributions are
completely known as: 
\begin{eqnarray}
\phi _{N} &=&-\frac{1}{\eta }\tau ^{5/8}~,  \label{phiN} \\
\phi _{1PN} &=&-\frac{1}{\eta }\left( \frac{3715}{8064}+\frac{55}{96}\eta
\right) \tau ^{3/8}~,  \label{phi1PN} \\
\phi _{1.5PN} &=&-\frac{3}{4\eta }\left( \frac{1}{4}\beta -\pi \right) \tau
^{1/4}~,  \label{phi1.5PN} \\
\phi _{2PN} &=&-\frac{1}{\eta }\left( \frac{9275495}{14450688}+\frac{284875}{%
258048}\eta +\frac{1855}{2048}\eta ^{2}-\frac{15}{64}\sigma \right) \tau
^{1/8}~,  \label{phi2PN} \\
\phi _{2.5PN} &=&-\frac{1}{\eta }\left[ \left( -\frac{38645}{172032}+\frac{65%
}{2048}\eta \right) \pi +\beta _{PN}\right] \ln \tau ~.  \label{phi2.5PN}
\end{eqnarray}%
Here the dimensionless time parameter $\tau =\eta (t_{c}-t)/5m$ is
decreasing until the merger. The contribution at 1PN is purely relativistic,
computed in \cite{phi1PN}. At 1.5 PN appear the leading-order tail
contributions \cite{phi1.5PNa}-\cite{phi1.5PNd}, together with the
spin-orbit contributions \cite{phiSOSS}, \cite{Kidder}, encompassed in the
coefficient 
\begin{equation}
\beta =\frac{1}{12}\sum_{i=1}^{2}\frac{S_{i}\cos \kappa _{i}}{m_{i}^{2}}%
\left( 113\frac{m_{i}^{2}}{m^{2}}+75\eta \right) \ .  \label{SO}
\end{equation}

At 2PN there are both relativistic contributions \cite{phi2PN} and finite
size effects, collected into the coefficient \label{spincoeffs} 
\begin{eqnarray}
\sigma &=&\sigma _{S_{1}S_{2}}+\sigma _{SS-self}+\sigma _{QM}\ ,  \label{all}
\\
\sigma _{S_{1}S_{2}} &=&\frac{S_{1}S_{2}}{48\eta m^{4}}(-247\cos \gamma
+721\cos \kappa _{1}\cos \kappa _{2})\ ,  \label{SS} \\
\sigma _{SS-self} &=&\frac{1}{96m^{2}}\sum_{i=1}^{2}\left( \frac{S_{i}}{m_{i}%
}\right) ^{2}\left( 6+\sin ^{2}\kappa _{i}\right) \ ,  \label{self} \\
\sigma _{QM} &=&-\frac{5}{2}\sum_{i=1}^{2}p_{i}\left( 3\cos ^{2}\kappa
_{i}-1\right) \ ,  \label{QM}
\end{eqnarray}%
The enlisted contributions originate in the spin-spin \cite{phiSOSS}, \cite%
{Kidder}, self-spin \cite{MVG} and quadrupole-monopole \cite{Poisson}
interactions.

\begin{figure}[tph]
\centering \resizebox{.6\columnwidth}{!} {\includegraphics[width=2cm]{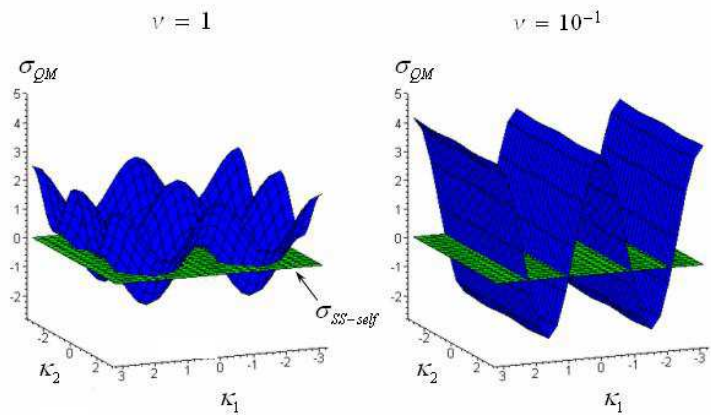}%
}
\caption{(Color online). The quadrupole-monopole contribution $\protect%
\sigma _{QM}$ (blue wavy surface) represented for equal masses and $\protect%
\nu =10^{-1}$ in the parameter space $\left( \protect\kappa _{1},\protect%
\kappa _{2}\right) $. The coefficient $\protect\sigma _{QM}$ strongly
depends on the spin configuration, in particular in the $\protect\nu %
=10^{-1} $ case only on the relative angle between the dominant spin and
orbital angular momentum. The $\protect\sigma _{QM}$ contribution increases
considerably with decreasing mass ratio. For comparison, the much smaller
self-spin contribution $\protect\sigma _{SS-self}$ is also represented
(green levelled surface). The values of $\protect\sigma _{SS-self}$ are
positive, but well below $1$. The plots are for maximally spinning binaries,
with the quadrupole moment induced by rotation.}
\label{fig1}
\end{figure}

The dependence of the spin coefficients $\sigma _{QM}$ and$~\sigma
_{SS-self} $ on the relative angles $\kappa _{i}$ is represented both for
equal masses and $\nu =10^{-1}$ on Fig. \ref{fig1}, under the assumptions of
maximal rotation and quadrupole moment generated purely by rotation.
Although formally both contributions are of 2PN order, the numerical values
of the coefficients involved imply that the quadrupole-monopole contribution
clearly dominates over the self-spin contribution, which only takes values
in the narrow interval $\sigma _{SS-self}\in \left( 1/32,~7/96\right)
\sum_{i=1,2}\left( m_{i}/m\right) ^{2}$.

The coefficient $\sigma _{QM}$ strongly depends on the spin configuration,
in particular in the $\nu =10^{-1}$ case only on the relative angle between
the dominant spin and orbital angular momentum. By comparing the two panels
of the figure, we see that the $\sigma _{QM}$ contribution increases
considerably with decreasing mass ratio. 

\begin{figure}[tph]
\centering \resizebox{1\columnwidth}{!} {%
\includegraphics[width=7cm]{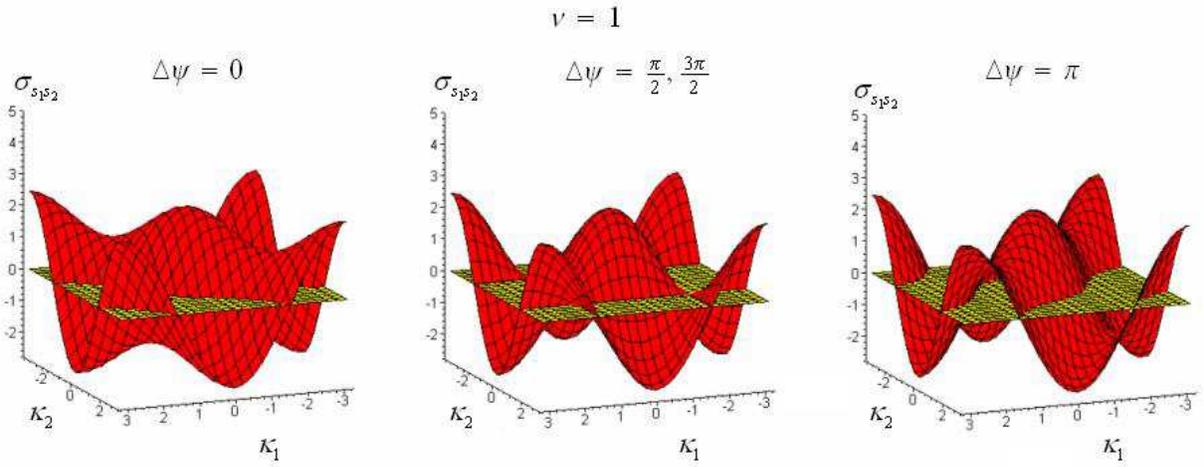}}
\caption{(Color online). The dependence of the spin-spin parameter $\protect%
\sigma _{S_{1}S_{2}}$ (red wavy surfaces) on the angles $\protect\kappa _{i}$
for equal masses and various values of the relativ spin azimuthal angle $%
\Delta \protect\psi =0,\protect\pi /2,\protect\pi ,3\protect\pi /2$. For
comparison the zero level surface is also represented. Plot for the maximal
rotation case.}
\label{fig2}
\end{figure}

\begin{figure}[tph]
\centering\resizebox{1\columnwidth}{!} {%
\includegraphics[width=7cm,]{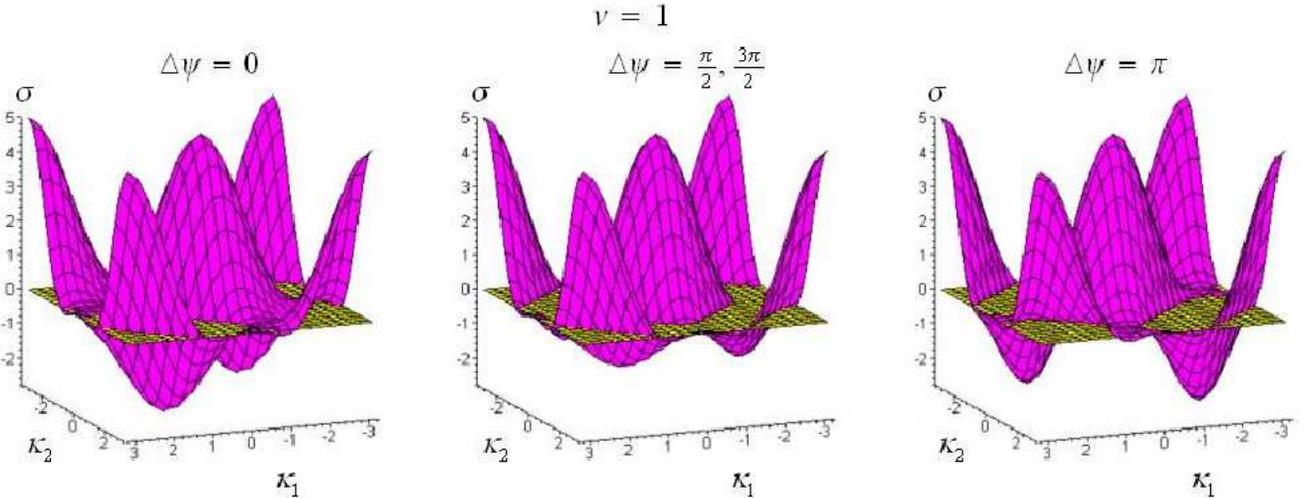}}
\caption{(Color online). Same as on Fig \protect\ref{fig2}, but for the
total coefficient $\protect\sigma $ (lilac wavy surfaces). Although there
are particular configurations (at the intersection with the zero level
surfaces) for which the various contributions to $\protect\sigma $ cancel
out each other, for the vaste majority of the parameter space $\protect%
\sigma \neq 0$.\thinspace $\ $The positive values are typically higher than
the negative ones due to the QM contribution.}
\label{fig3}
\end{figure}

In the \textit{equal mass case} we also show the dependence of $\sigma
_{S_{1}S_{2}}$ and of the total $\sigma $ on the parameter space $\left(
\kappa _{1},~\kappa _{2}\right) $ for various relative spin azimuthal angles 
$\Delta \psi $ (Figs. \ref{fig2} and \ref{fig3}). Here $\Delta \psi =\psi
_{2}-\psi _{1}$ is the difference between the azimuthal angles $\psi _{i}$
of the spin vectors. 
As the vertical scale for all these figures is the same, we can clearly see
that the quadrupole-monopole contribution dominates for a huge part of the
parameter space. Also, in general $\sigma \neq 0$ and its positive values
are higher than the negative ones due to the QM contribution. However there
is a small set of particular configurations for which the contributions to $%
\sigma $ cancel out each other.

The coefficients $\sigma _{S_{1}S_{2}}$ and $\sigma $ in the \textit{%
non-equal mass case} will be discussed in the remaining Sections of the
paper. They are more complicated because (anticipating the results of the
next Section) the relative spin azimuthal angle $\Delta \psi $ becomes
time-dependent for $\nu \neq 1$ due to precessions.

Finally at 2.5 PN there are contributions to $\phi $ from the first PN
correction to SO \cite{PNSO}, \cite{PNSO2}, expressed by%
\begin{equation}
\beta _{PN}=\sum_{i=1}^{2}\frac{S_{i}\cos \kappa _{i}}{m_{i}^{2}}\left[
\left( \frac{965}{3584}\eta +\frac{681\,145}{516\,096}\right) \frac{m_{i}^{2}%
}{m^{2}}+\left( \frac{37265}{57344}+\frac{1735}{7168}\eta \right) \,\eta %
\right] \ ,  \label{PNSO}
\end{equation}%
and tail effects \cite{phi2.5PN}. We note that the logarithmic terms arising
from the latter can be eliminated through a modified definition of the phase 
\cite{BFIS}

The list of contributions known at 3PN and 3.5PN are not exhaustive. At 3PN
there are known relativistic and tail contributions \cite{phi3PN3.5PN} 
\begin{eqnarray}
\phi _{3PN} &=&-\frac{1}{\eta }\biggl\{\frac{831032450749357}{57682522275840}%
-\frac{53}{40}\pi ^{2}-\frac{107}{56}C  \notag \\
&&+\frac{107}{448}\ln \left( \frac{\tau }{256}\right) +\left[ -\frac{%
126510089885}{4161798144}+\frac{2255}{2048}\pi ^{2}\right] \eta  \notag \\
&&+\frac{154565}{1835008}\eta ^{2}-\frac{1179625}{1769472}\eta ^{3}\biggr\}%
\tau ^{-1/8}\ ,  \label{phi3PN}
\end{eqnarray}%
where $C=\lim_{n\rightarrow \infty }\left[ \sum_{k=1}^{n}\left( 1/k\right)
-\ln n\right] \approx 0.577$ is the Euler-Mascheroni constant\footnote{%
The Euler-Mascheroni constant appears, among other places, in expressions
involving the exponential integral, the Laplace transform of the natural
logarithm, an inequality for Euler's totient function and solution of the
second kind to Bessel's equation \cite{WW}.}. However the contributions from
the PN corrections of the SS, self and QM contributions are not known.

At 3.5PN only the tail contribution has been calculated \cite{phi3PN3.5PN}

\begin{equation}
\phi _{3.5PN}=-\frac{1}{\eta }\left( \frac{188516689}{173408256}+\frac{488825%
}{516096}\eta -\frac{141769}{516096}\eta ^{2}\right) \pi \tau ^{-1/4}~,
\label{phi3.5PN}
\end{equation}%
but clearly there are numerous other contributions, as the 2PN correction to
the SO contribution, SO-SS, SO-QM couplings etc.

It is the main purpose of this paper to add new spin-spin contributions at
3PN, arising from the spin-orbit precession of the spin vectors and discuss
their implication to the definition of an effective $\sigma $ at 2PN.

\section{Leading order precessional contribution to $\protect\sigma $}

The spin vectors undergo a complicated quasi-precessional motion \cite{BOC},
and the angles $\kappa _{i}$ and $\gamma $ evolve accordingly. In Ref. \cite%
{GPV} it was derived that the leading order contributions to the
quasi-precessional variation of the angles $\kappa _{i}$ is of $\varepsilon
^{3/2}$ order, while $\gamma $ evolves in the $\varepsilon $ order.
Supplementing the SO and SS contributions already considered in Ref. \cite%
{GPV} with the QM contributions (with the quadrupole moment arising from
pure rotation), by employing Eqs. (\ref{p1})-(\ref{S1L}) and the method
described in a footnote of Ref. \cite{spinspin2}, according to which $%
O(\delta x)=O(\dot{x})r\varepsilon ^{-1/2}$, we have checked that these
estimates are not changed, although the leading order changes in $\kappa
_{i} $ acquire QM type contributions. However the leading order change in $%
\gamma $ arises solely from the SO precession.

With the angles $\kappa _{i}$ and $\gamma $ non-constant, the coefficients $%
\beta ,~\sigma $ and $\beta _{PN}$ also evolve in time. From among the
relative spin angles the contributions $\sigma _{QM}$ and $\sigma _{SS-self}$
contain only $\kappa _{i}$. As these start to vary at $\varepsilon ^{3/2}$
order, the variations of $\kappa _{i}$ would contribute only at 3.5PN
orders. The same is true for the $\kappa _{i}$-dependence of $\sigma
_{S_{1}S_{2}}.$ However the variation at $\varepsilon $ order of $\gamma $
induces 3PN corrections to $\sigma _{S_{1}S_{2}}$. Thus the leading order
precessional contribution to $\sigma $ is caused by the leading order
precessional evolution of the spin angle $\gamma $, of SO origin.

Therefore in what follows we derive the time dependence of $\gamma $. From
the SO precession of the spin vectors (the SO contribution in Eq. (2.17c) of
Ref. \cite{GPV}, with the change of notation $\eta \rightarrow \nu $) we
find 
\begin{equation}
(\cos \mathbf{\skew{25}{\dot}{\gamma}}\mathbf{)\!}=-\!\frac{3L\left( \nu
-\nu ^{-1}\right) }{2r^{3}}\sin \kappa _{1}\sin \kappa _{2}\sin \Delta \psi
\!\ .  \label{gamma2}
\end{equation}%
By taking the time derivative of the spherical cosine identity%
\begin{equation}
\cos \gamma =\cos \kappa _{1}\cos \kappa _{2}+\cos \Delta \psi \sin \kappa
_{1}\sin \kappa _{2}\ .  \label{gombi}
\end{equation}%
we get%
\begin{eqnarray}
\sin \gamma ~\frac{d\gamma }{dt} &=&\sin \kappa _{1}\sin \kappa _{2}\sin
\Delta \psi \frac{d}{dt}\Delta \psi  \notag \\
&&+\cos \Delta \psi \left[ \cos \kappa _{1}\frac{d}{dt}\left( \cos \kappa
_{2}\right) +\cos \kappa _{2}\frac{d}{dt}\left( \cos \kappa _{1}\right) %
\right] \!\ .  \label{gombi_dot}
\end{eqnarray}%
The second term can be dropped as we need to consider this relation only to $%
\varepsilon $ accuracy. A comparison of Eqs. (\ref{gamma2}) and (\ref%
{gombi_dot}) gives:

\begin{equation}
\frac{d}{dt}\Delta \psi =\frac{3L_{N}\left( \nu ^{-1}-\nu \right) }{2r^{3}}~,
\label{Depsi}
\end{equation}%
whenever $\kappa _{i},\Delta \psi \neq 0$. For equal masses $\Delta \psi =$%
\textit{const}, therefore $\gamma $ is also unchanged by precessional
effects to $\varepsilon $ order. Therefore $\sigma _{S_{1}S_{2}}$ can be
considered constant even at 3PN for $\nu =1$.

For any $\nu \neq 1$ Eq. (\ref{Depsi}) can be integrated in the following
way. First we note that in order to calculate the leading order change in $%
\Delta \psi $, all quantities are needed only to Keplerian order. We also
specialize to circular orbits $r=a$, as required by the derived expression
of the phase. We then pass to integration in terms of the true anomaly $\chi 
$, defined as%
\begin{equation}
\frac{d\chi }{dt}=\frac{L}{\mu a^{2}}\ ,
\end{equation}%
and employ the Kepler equation to establish the connection between the time
and the eccentric anomaly $\xi $ (which to leading order and circular orbits
coincides with the true anomaly $\chi $):%
\begin{equation}
n\left( t-t_{0}\right) =\chi ~,  \label{Kepler}
\end{equation}%
where $n=2\pi /T_{orbit}$. By choosing $t_{0}=0$, we obtain the required
explicit time dependence of $\gamma $:%
\begin{equation}
\cos \gamma =\cos \kappa _{1}\cos \kappa _{2}+\cos \left[ \left( \Delta \psi
\right) _{0}+\frac{3\mu n\left( \nu ^{-1}-\nu \right) }{2a}t\right] \sin
\kappa _{1}\sin \kappa _{2}~,  \label{gamma_solve}
\end{equation}%
which gives a new time-dependent 3PN spin contribution to $\sigma
_{S_{1}S_{2}}$.

The angle $\gamma $ (and the respective 3PN contribution to $\sigma
_{S_{1}S_{2}}$) has a periodicity with period%
\begin{equation}
T_{3PNSS}=\frac{4\pi a}{3\mu \left( \nu ^{-1}-\nu \right) n}=\frac{2a}{3\mu
\left( \nu ^{-1}-\nu \right) }T_{orbit}~.
\end{equation}%
With the approximation valid for quasicircular orbits: $T_{orbit}=2T_{wave}$%
, also employing $\varepsilon \approx m/a$ and $\eta ^{-1}=m/\mu =\left(
2+\nu +\nu ^{-1}\right) $, we get:%
\begin{equation}
\frac{T_{3PNSS}}{T_{wave}}=\frac{4}{3\left( \nu ^{-1}-\nu \right)
\varepsilon \eta }=\frac{4\left( 2+\nu +\nu ^{-1}\right) }{3\left( \nu
^{-1}-\nu \right) \varepsilon }~.
\end{equation}%
It has been shown in \cite{spinflip}, that the typical mass ratio at
galactic black hole encounters is between $1/3$ to $1/30$, such that the
mass ratio $\nu =10^{-1}$ can be considered fairly representative. We adopt
this value in what follows. Then%
\begin{equation}
\frac{T_{3PNSS}}{T_{wave}}\approx 1.\,\allowbreak 63~\varepsilon ^{-1}~.
\end{equation}%
We have therefore found that for the representative mass ratio the
time-scale of variation of the angle $\gamma $, (and the related change in $%
\sigma _{S_{1}S_{2}}$) is of $\varepsilon ^{-1}$ times larger order, than
the period of gravitational waves.

\section{The time-evolving $\protect\sigma _{S_{1}S_{2}}$ and $\protect%
\sigma $ in the unequal mass case $\protect\nu =10^{-1}$}

We remark on Eq. (\ref{gamma_solve}) that on a time-scale $T_{3PNSS}$, the
time-dependent contribution averages out:%
\begin{equation}
\overline{\cos \gamma }=\cos \kappa _{1}\cos \kappa _{2}~,
\end{equation}%
therefore the average of the spin coefficient $\sigma _{S_{1}S_{2}}$ becomes%
\begin{equation}
\overline{\sigma _{S_{1}S_{2}}}=\frac{79S_{1}S_{2}}{8\eta m^{4}}\cos \kappa
_{1}\cos \kappa _{2}\ .  \label{SSrenorm}
\end{equation}%
We call this the \textit{renormalized} spin-spin coefficient in the phase of
gravitational waves. A comparison with the standard expression (\ref{SS})
shows that the renormalized expression is simpler than the original one.

The spin coefficient $\sigma _{S_{1}S_{2}}$ to 3PN accuracy can be then
rewritten as a sum of the renormalized coefficient and a time-varying part:%
\begin{eqnarray}
\sigma _{S_{1}S_{2}} &=&\overline{\sigma _{S_{1}S_{2}}}+\delta \sigma ~,
\label{si} \\
\delta \sigma &=&-\frac{247S_{1}S_{2}}{48\eta m^{4}}\sin \kappa _{1}\sin
\kappa _{2}\cos \left[ \left( \Delta \psi \right) _{0}+\frac{3\mu n\left(
\nu ^{-1}-\nu \right) }{2a}t\right] \ .  \label{desi}
\end{eqnarray}%
The second term is periodic (with period $T_{3PNSS}\approx \varepsilon
^{-1}T_{wave}$ for the typical binaries with mass ratio $\nu =10^{-1}$), and
as such, it averages out over a time scale $\varepsilon ^{-1}T_{wave}$.

\begin{figure}[tph]
\centering \resizebox{1\columnwidth}{!} 
{\includegraphics[width=7cm,]{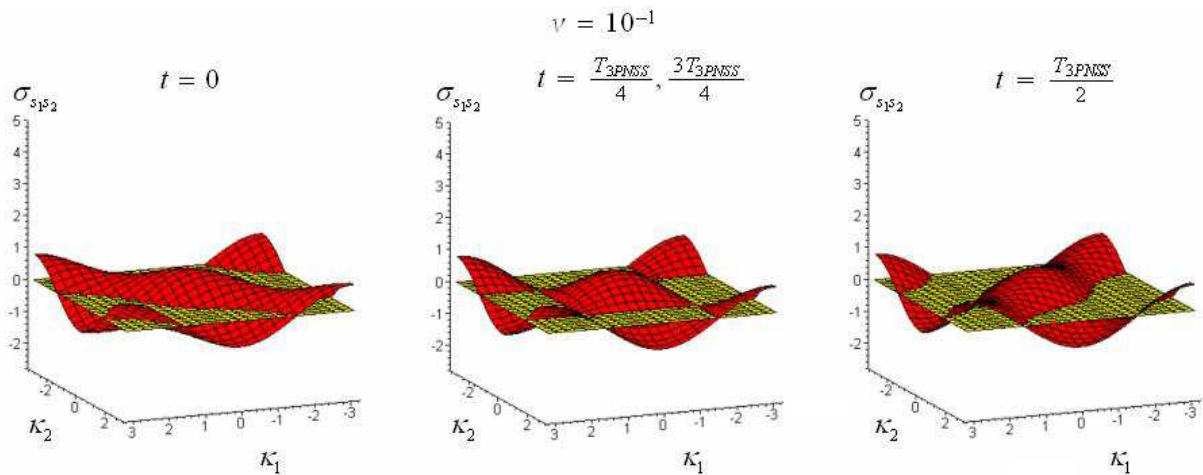}}
\caption{(Color online). The time-evolving spin-spin parameter $\protect%
\sigma _{S_{1}S_{2}}$ for mass ratio $\protect\nu =10^{-1}$ represented
every quarter period during $T_{3PNSS}$.}
\label{fig4}
\end{figure}

\begin{figure}[tph]
\centering\resizebox{1\columnwidth}{!} 
{\includegraphics[width=7cm,]{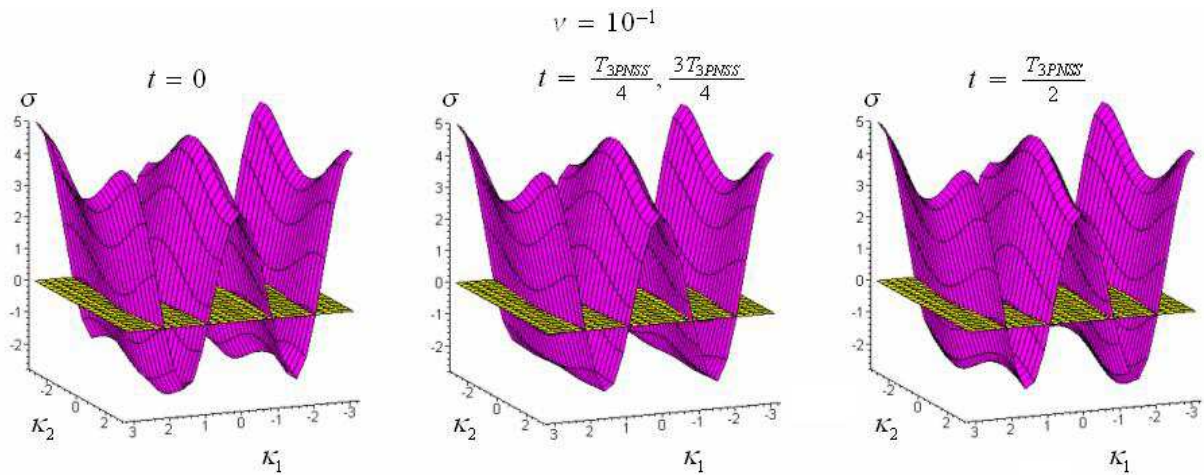}}
\caption{(Color online). The time-evolving parameter $\protect\sigma $ for
mass ratio $\protect\nu =10^{-1}$ represented every quarter period during $%
T_{3PNSS}$. One can see that the dominant contribution is from the
quadrupole-monopole interaction.}
\label{fig5}
\end{figure}

On Figs \ref{fig4} and \ref{fig5} we have represented the time-evolving
parameters $\sigma _{S_{1}S_{2}}$ and $\sigma $ for the mass ratio $\nu
=10^{-1}$ in a time sequence sampled every quarter period during $T_{3PNSS}$%
. \footnote{%
Animated gif-s showing the temporal evolution of both $\sigma _{S_{1}S_{2}}$
and $\sigma $ are available at \cite{webpage}.} 
While the importance of the spin-spin contribution decreased as compared to
the equal mass case (the scales on Figs \ref{fig2} and \ref{fig4} are the
same for easy comparison), the quadrupole-monopole contribution is enhanced
by a small $\nu $ (Fig \ref{fig1}). As a result, for LISA sources the
dominant contribution comes from the quadrupole-monopole interaction and the
time-varying spin-spin contribution only induces a wobbling of the crests
determined by the quadrupole-monopole interaction (compare Figs \ref{fig1}
and \ref{fig5}).

\section{Analytical and numerical contributions to the accumulated number of
gravitational wave cycles}

The phase of the gravitational waves (\ref{phase}) emerges by two
integrations over time from the expression of the time derivative of the
orbital angular frequency $\dot{\omega}$, see Ref. \cite{MVG}. The spin
coefficient $\sigma _{S_{1}S_{2}}$ first appears in the latter expression
and to a 2PN accuracy it can be considered constant, hence it also emerges
in the phase.

To a higher accuracy the coefficient acquires time-dependence from the
precession equations, which should change the results of the integration.
Alternatively, the contribution of the leading order SO precession can be
taken into account by the averaging procedure described in the previous
section. Then the spin coefficient $\overline{\sigma _{S_{1}S_{2}}}$,
regarded as a constant, formally leads to the same expression of the phase
and accumulated number of gravitational wave cycles $\mathcal{N}=\left( \phi
_{c}-\phi \right) /\pi $ as Eq. (12) of Ref. \cite{MVG}, with the
renormalized coefficient $\overline{\sigma _{S_{1}S_{2}}}$ replacing $\sigma
_{S_{1}S_{2}}$. Obviously this result is approximate rather then exact. The
question comes, how good is the approximation?

As the exact time integrations turn out to be cumbersome and the result is
not expressible in terms of elementary functions, in order to answer this
question we calculate numerically the corresponding contribution to the
accumulated number of gravitational wave cycles. We proceed in the following
way. Eq. (7) of Ref. \cite{MVG} gives the evolution of the orbital angular
frequency $\omega $ in time to 2PN accuracy. By inserting the time dependent
expression (\ref{si}) into this equation, and keeping only the Newtonian and
the time-dependent correction of the spin-spin terms, we obtain:%
\begin{equation}
\frac{d\left( m\omega \right) }{d\tau }=-96\left( m\omega \right)
^{11/3}-96\left( m\omega \right) ^{5}\delta \sigma \ .  \label{omegadot}
\end{equation}%
The solution of this equation can be also split into the known Newtonian
contribution and a term of $\delta \sigma $ order:%
\begin{equation}
m\omega (\tau )=\frac{\tau ^{-3/8}}{8}+m\omega _{\delta \sigma }\left( \tau
\right) \ .  \label{omegatau}
\end{equation}%
Here the dimensionless time $\tau $ is defined as $\tau =\eta \left(
t_{c}-t\right) /5m$, with $\left( t-t_{c}\right) $ the time until the
coalescence. (Note that we are discussing only the precessional corrections
to the spin-spin contribution in this paper, without attempting to encompass
other cumbersome post-Newtonian corrections of the spin-spin effects arising
from the dynamics.)

Rather than giving the number of cycles left in terms of $\tau $, we express
it as an integral over the orbital angular frequency%
\begin{equation}
\mathcal{N}=\mathcal{N}_{N}+\mathcal{N}_{\delta \sigma }=\frac{1}{\pi }%
\int_{\tau _{i}}^{\tau _{f}}\omega \left( \tau \right) d\tau =\frac{1}{\pi }%
\int_{\omega _{i}}^{\omega _{f}}\frac{\omega }{\dot{\omega}}d\omega ~.
\label{N}
\end{equation}%
Here $\dot{\omega}$ is given by Eq. (\ref{omegadot}). In order to express $%
\delta \sigma \left( \omega \right) $, from Eq. (\ref{desi}) by employing
Kepler's third law $\omega ^{2}a^{3}=m$ and $n=\omega $ (valid to leading
order), we obtain%
\begin{equation}
T_{3PNSS}=\frac{4\pi m}{3\eta \left( \nu ^{-1}-\nu \right) \left( m\omega
\right) ^{5/3}}~.
\end{equation}%
We also express $t$ as function of $\omega $ to leading order from Eq. (\ref%
{omegatau}) as 
\begin{equation}
t-t_{c}=-\frac{5m}{2^{8}\eta }\left( m\omega \right) ^{-8/3}~.
\end{equation}%
Thus%
\begin{equation}
\delta \sigma =-\frac{247S_{1}S_{2}}{48\eta m^{4}}\sin \kappa _{1}\sin
\kappa _{2}\cos \left[ \left( \Delta \psi \right) _{0}+\frac{3\eta \left(
\nu ^{-1}-\nu \right) t_{c}}{2m}\left( m\omega \right) ^{5/3}-\frac{15\left(
\nu ^{-1}-\nu \right) }{2^{9}}\left( m\omega \right) ^{-1}\right] ~.
\label{desiom}
\end{equation}%
Inserting the expression (\ref{omegadot}), with $\delta \sigma $ given by
Eq. (\ref{desiom}) into Eq. (\ref{N}) we obtain the number of cycles in the
frequency range $\left( \omega _{i},\omega _{f}\right) $. This enables us to
give accurate numerical results for the accumulated number of cycles.

For comparison we select three LISA sources with unequal mass given in
Tables I and II of Ref. \cite{BBW} and we enlist in Table \ref{Table1} the
corresponding spin-spin contributions calculated for particular (constant)
values of the angles $\kappa _{i}$ in three different ways: (a) by employing
Eq. (10) of Ref. \cite{MVG}; (b) in the same way, but with our renormalized
spin-spin coefficient $\overline{\sigma _{S_{1}S_{2}}}$ replacing $\sigma
_{S_{1}S_{2}}$; (c) employing a numerical integration based on the exact
method described in this section. For comparison we also give the
corresponding Newtonian, PN and QM contributions. The data confirms that for
these non-equal mass binaries the QM contribution dominates over the
spin-spin contribution. It is also immediate to see that the renormalized
spin-spin coefficient leads to much closer values to the results of the
numerical integration, then the 2PN-accurate coefficient. 
\begin{table}[t]
\caption{Newtonian, PN, QM and spin-spin contributions to the accumulated
number of gravitational wave cycles for three LISA sources with unequal
mass, for particular values of the angles $\protect\kappa _{i}$ and $\left(
\Delta \protect\psi \right) _{0}$. The relevant frequency ranges (with $f=%
\protect\omega /2\protect\pi $) are also indicated. The spin-spin
contributions are computed in three distinct ways (a) in a 2PN rigurous
formalism ($\mathcal{N}_{\protect\sigma _{SS}}$); (b) with the renormalized
spin-spin coefficient ($\mathcal{N}_{\bar{\protect\sigma}_{SS}}$); and (c)
taking into account the exact precessional evolutions up to 3PN by numerical
integration ($\mathcal{N}_{\bar{\protect\sigma}_{SS}+\protect\delta \protect%
\sigma _{SS}(t)}$). The numbers obtained with the renormalized coefficient
are within $0.02$ cycles agreement with the numerical data, presenting an
improvement up to $0.6$ cycles over the 2PN accurate expression. The numbers
in the table confirm that for unequal masses the QM contribution dominates
over the spin-spin contribution.}
\label{Table1}
\begin{center}
\begin{tabular}{|l|c|c|c|}
\hline
$%
\begin{array}{c}
\left( \Delta \psi \right) _{0}=15^{\circ } \\ 
\kappa _{1}=\kappa _{2}=30^{\circ }%
\end{array}%
$ & $%
\begin{array}{c}
1.4M_{\odot } \\ 
400M_{\odot }%
\end{array}%
$ & $%
\begin{array}{c}
10^{4}M_{\odot } \\ 
10^{5}M_{\odot }%
\end{array}%
$ & $%
\begin{array}{c}
10^{6}M_{\odot } \\ 
10^{7}M_{\odot }%
\end{array}%
$ \\ \hline\hline
\multicolumn{1}{|c|}{$f_{in}(Hz)$} & $4.601\times 10^{-2}$ & $4.199\times
10^{-4}$ & $2.361\times 10^{-5}$ \\ 
\multicolumn{1}{|c|}{$f_{fin}(Hz)$} & $1.000$ & $3.997\times 10^{-2}$ & $%
3.997\times 10^{-4}$ \\ \hline\hline
\multicolumn{1}{|c|}{$\mathcal{N}_{N}$} & $2294624.602$ & $21055.705$ & $%
1174.257$ \\ \hline
\multicolumn{1}{|c|}{$\mathcal{N}_{PN}$} & $35363.312$ & $677.247$ & $%
114.536 $ \\ \hline
\multicolumn{1}{|c|}{$\mathcal{N}_{\sigma _{QM}}$} & $-432.606$ & $-13.712$
& $-6.028$ \\ \hline\hline
\multicolumn{1}{|c|}{$\mathcal{N}_{\sigma _{SS}}$} & $-2.986$ & $-2.678$ & $%
-1.177$ \\ \hline
\multicolumn{1}{|c|}{$\mathcal{N}_{\bar{\sigma}_{SS}}$} & $-3.588$ & $-3.218$
& $-1.415 $ \\ \hline
\multicolumn{1}{|c|}{$\mathcal{N}_{\bar{\sigma}_{SS}+\delta \sigma _{SS}(t)}$%
} & $-3.588$ & $-3.207$ & $-1.404$ \\ \hline
\end{tabular}%
\end{center}
\end{table}

\section{Concluding Remarks}

We have discussed the contributions induced by the spin-spin and
quadrupole-monopole coupling to the accumulated orbital phase. We have shown
that the contributions from the self-coupling of the spins are much smaller
than the QM and SS contributions in all cases (Fig \ref{fig1}), therefore
they can be dropped at a 2PN accuracy. The QM contributions depend on the
parameter space $\left( \kappa _{1},~\kappa _{2}\right) $, while the SS
contributions on $\left( \kappa _{1},~\kappa _{2},\text{ }\Delta \psi
\right) $ [or alternatively on $\left( \kappa _{1},~\kappa _{2},\text{ }%
\gamma \right) $], and were represented on these parameter spaces.

For equal mass compact binaries we have shown that the QM contributions
(left panel of Fig \ref{fig1}) and the SS contributions (Fig \ref{fig2}) are
of comparable size, although the SS contributions are slightly higher. These
two contributions (at least for positive values) happen to roughly amplify
each other, thus their combined contribution (Fig \ref{fig3}) has
approximately twice as large positive amplitude, than the individual
contributions. There is also a set of measure zero in the parameter space,
for which the total contribution vanishes.

For typical LISA sources with unequal masses $\nu =10^{-1}$ the SS
contribution becomes time-dependent due to the variation of the angle $%
\gamma $ by SO precession. The time-evolution of the SS coefficient was
shown on Fig \ref{fig4}. By comparing with the QM contribution for $\nu
=10^{-1}$ (right panel of Fig \ref{fig1}), we see that the SS contribution
is much smaller. As consequence, when we add these contributions, the
combined coefficient shows the pattern of the QM contribution, modulated in
time by the SS contribution (Fig \ref{fig5}).

The time variation of the SS component was shown to be periodic, with a
period $\varepsilon ^{-1}$ times longer than the period of the gravitational
wave. By separating from the SS contribution a time-varying component which
averages to zero on this time-scale, we may call the rest as a renormalized
SS coefficient $\overline{\sigma _{S_{1}S_{2}}}$. Besides of being constant
up to 3PN, it has a simpler expression, than the previously known 2PN SS
coefficient $\sigma _{S_{1}S_{2}}$.

Therefore we find useful to introduce in the unequal mass case the total
renormalized coefficient $\overline{\sigma }=\sigma _{QM}+\overline{\sigma
_{S_{1}S_{2}}}+\sigma _{SS-self}$ for maximally rotating black holes with
quadrupole moment due to rotation. It has the expression 
\begin{equation}
\overline{\sigma }\approx \frac{5}{2}\left[ 3\cos ^{2}\kappa _{1}-1+\nu
\left( 2+\frac{79}{20}\cos \kappa _{1}\cos \kappa _{2}-6\cos ^{2}\kappa
_{1}\right) \right] ~,
\end{equation}%
which was obtained by keeping only the leading and first order QM and SS
contributions in $\nu $. The renormalized $\overline{\sigma }$ is constant
in time up to 3PN and as such is more convenient to use in wave detection,
than $\sigma $, which is constant only to 2PN. For all numerical examples
studied, the renormalized coefficient gave results within\ $0.02$ cycle
accuracy as compared to the numerical results. This represented an
improvement of $0.4\div 0.6$ cycles over the previously employed 2PN
accurate coefficient.

\begin{figure}[tph]
\centering{\includegraphics[width=7.5cm,]{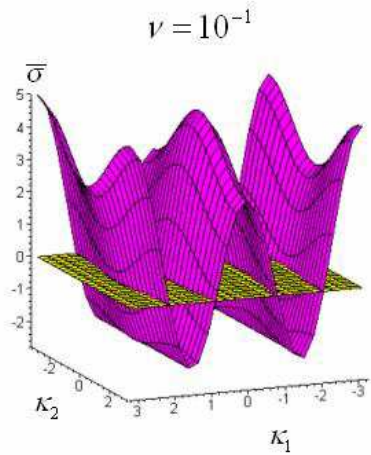}}
\caption{(Color online). The renormalized parameter $\bar{\protect\sigma}$
for mass ratio $\protect\nu =10^{-1}$. The dominant contribution comes from
the quadrupole-monopole interaction (compare with the second panel of Fig 
\protect\ref{fig1}), which is slightly modulated by the renormalized
spin-spin contribution.}
\label{fig6}
\end{figure}

The dependence of $\overline{\sigma }$ on the parameter space $\left( \kappa
_{1},~\kappa _{2}\right) $ is shown on Fig \ref{fig6}. It is very similar to
the pure quadrupole-monopole effect, represented in the second panel of Fig %
\ref{fig1}, which is slightly modulated by the renormalized spin-spin
contribution. 

\textit{Due to the arguments presented in this paper we propose to replace }$%
\sigma $\textit{\ by }$\overline{\sigma }$\textit{\ in the expression of the
gravitational wave phase}, Eq. (\ref{phi2PN}), \textit{whenever it is
applied for LISA sources}.

In a similar way to the time dependence and renormalization of $\sigma $, a
time dependence and renormalization of the spin-orbit contribution $\beta $
can be derived, the calculation of which involves to integrate the combined
SO, SS and QM quasi-precessional time evolution of the angles $\kappa _{i}$
at $\varepsilon ^{3/2}$ order. This has been done for the equal mass case 
\cite{Racine}, however it seems to be far from obvious for $\nu \neq 1$.
Therefore we defer the study of this effect to a future work.

\section{Acknowledgements}

L\'{A}G was succesively supported by the the Hungarian Scientific Research
Fund (OTKA) grant no. 69036, the J\'{a}nos Bolyai Grant of the Hungarian
Academy of Sciences, the London South Bank University Research Opportunities
Fund and the Pol\'{a}nyi Program of the Hungarian National Office for
Research and Technology (NKTH). BM was supported by OTKA grants no. 68228
and 69036.

\end{document}